\documentclass{article}
\usepackage{graphicx} 
\usepackage{amsthm, amsmath, amsfonts,mathtools, fullpage, graphicx, wrapfig, tikz, caption, subcaption}
\usepackage{natbib}
\usepackage{amssymb}
\usepackage{verbatim}
\usepackage{relsize}
\usepackage{listings}
\usepackage{accents}
\usepackage{authblk}
\usepackage{xcolor}
\usepackage{hyperref}
\usepackage{graphicx}
\usepackage{multirow}

\newtheorem{assumption}{Assumption}

\newtheorem{remark}{Remark}
\title{Structural Nested Mean Models Under Parallel Trends with Interference}
\author[1]{Zach Shahn}
\author[2]{Paul N Zivich}
\author[3]{Audrey Renson}
\affil[1]{CUNY Graduate School of Public Health and Health Policy}
\affil[2]{Department of Epidemiology, Gillings School of Global Public Health, UNC Chapel Hill}
\affil[3]{NYU Grossman School of Medicine}
\begin{document}

\maketitle

\section{Introduction}

Difference-in-differences (DiD) is a commonly used approach to assess effects of interventions, policies, or other exposures. In settings where DiD is applied, an exposure in one unit often impacts outcomes in other units, a phenomenon called `interference' (or, alternately, `spillover') in the causal inference literature \citep{hudgens2008toward}. For example, if one county's minimum wage increases, neighboring counties may also experience economic effects. As another example, housing vouchers offered to some families in a housing project may influence whether other families in the same housing project move \citep{sobel2006randomized}.

Interference is often taxonomized into `cluster' and `network' structures. In a cluster interference structure, the population is partitioned into clusters such that spillover effects may occur between any two units in the same cluster, but do not leave the cluster. Examples of clusters on which it might sometimes be reasonable to assume a cluster interference structure include households (with people as units) or schools (with classes as units). \citet{he2015structural} provided a simple ophthalmological example where individual units were eyes, and each cluster comprised both eyes of a subject. In network interference, spillover effects between units depend on their connection to each other on a network or graph. A simple example of network interference is spillover effects due to geographic proximity, where units are connected to their neighbors in physical space. Note that spillover effects from geographic proximity would not have a cluster interference structure because sets of neighbors do not form a partition.

Despite the occurrence of interference in DiD applications, standard DiD methods rely on an assumption that interference is absent, and comparatively little work has considered how to accommodate and learn about spillover effects within a DiD framework. \cite{clarke2017estimating} relies on parametric and effect homogeneity assumptions to identify network spillover effects of speed limits on traffic accidents in neighboring counties. \cite{hettinger2023estimation} relax these prior parametric assumptions for the point exposure setting and study network spillover effects of a beverage tax. \citet{fiorini2024simple} consider the time varying setting. They extend the Mundlak estimator of \citet{wooldridge2021two}, one of a host of methods for estimating effects of initial exposure under `staggered adoption' designs (\citep{roth2023,chaisemartin_review}), to accommodate and estimate spillover effects. However, due to limitations of the staggered DiD methods, \citet{fiorini2024simple} are forced to make the dubious simplifying assumption that spillover effects are zero in any units after receiving direct exposure. 

It is limiting to base estimation of time-varying effects in the presence of interference on staggered DiD approaches. \citet{shahn2022structural} enable estimation of the effects of repeated discrete, continuous, or mixed exposures via structural nested mean models (SNMMs) \citep{jamie1994,jamie1997,jamie2004,snmm_review2014} under a time-varying parallel trends assumption conditional on evolving exposure and covariate history.  Here, we extend the `DiD-SNMMs' of \citet{shahn2022structural} to accommodate interference in a time-varying DiD setting. Doing so enables estimation of a richer set of effects than previous DiD approaches.
For example, the proposed DiD-SNMMs can model how effects of direct or indirect exposures depend on past and concurrent (direct or indirect) exposure history. Even if interest does not directly center on spillover effects, it is still necessary to properly account for them when they are present to estimate other quantities, such as expected outcomes had the entire population been untreated. 

We consider both cluster and network interference structures. For cluster interference structures, we follow \citet{he2015structural}, who considered SNMMs under cluster interference and a sequential exchangeability assumption, in considering effects of full cluster exposure assignments on the vector of outcomes within a cluster. In the case of network interference, we adopt the exposure mapping formulation of interference of  \citet{aronow2017estimating,manski2013identification}. This formulation requires the analyst to specify a mapping that, for each unit, characterizes the aspects of the full sample exposure assignment that might affect the unit's outcome. An example of an exposure mapping would be to specify that counties' outcomes depend on their neighbors' exposure assignments via an indicator of whether any neighbor was treated. An alternative specification would be to specify that county outcomes depend on the number of treated neighbors. We formalize and further discuss exposure mappings later. 

The organization of the paper is as follows. In Section \ref{cluster}, we discuss identification and estimation of SNMMs for joint direct and spillover effects under parallel trends assumptions and cluster interference. In the subsection on identification, we discuss some causal estimands of interest that are byproducts of SNMM parameters. In Section \ref{network}, we do the same for network interference. In Section \ref{simulations}, we provide simulations illustrating the ability of SNMMs to estimate a variety of interesting causal contrasts in both interference settings. In Section \ref{sec:app}, we apply our methods to estimate direct and spillover effects of Medicaid expansion on uninsurance rates. In section \ref{sec:conclusion}, we conclude with some areas for future work. 

\section{Cluster Interference}\label{cluster}
\subsection{Notation and Effects of Interest}
Suppose that data are collected on $N$ clusters with $J$ units observed in each cluster at time points $1,\ldots,\tau$. We will denote random variables with capital letters and realizations of random variables with lower case letters. Let $A_{i,m,j}$ denote exposure for unit $j$ at time $m$ in cluster $i$, $L_{i,m,j}$ denote covariates for unit $j$ at time $m$ in cluster $i$ (possibly including cluster level covariates shared across units within a cluster), and $Y_{i,m,j}$ denote the outcome for unit $j$ at time $m$ in cluster $i$. Let underscoring by tilde, e.g., $\undertilde{Y}_{i,m}$, denote vectors of variables over units in a cluster. For any time varying variable $X$, let $\bar{X}_k$ denote $(X_1,\ldots,X_k)$, and let $\underline{X}_k$ denote $(X_k,\ldots,X_{\tau})$. Let $Y_{i,k,j}(\bar{\undertilde{a}}_{i,k-1})$ denote the potential outcome \citet{robins1986new} of unit $j$ in cluster $i$ at time $k$ under cluster exposure regime $\bar{\undertilde{a}}_{i,k-1}$. Accordingly, let $\undertilde{Y}_{i,k}(\bar{\undertilde{a}}_{i,k-1})$ denote the vector of potential outcomes of cluster $i$ at time $k$ under cluster exposure regime $\bar{\undertilde{a}}_{i,k-1}$. Hereafter, we  omit cluster-level subscripts when it does not lead to confusion since we assume data from different clusters are independent and identically distributed. 

\citet{he2015structural} define vector valued causal contrasts of interest for cluster level exposure assignments as
\begin{equation}
\undertilde{\gamma}^*_{m,k}(\bar{\undertilde{a}}_m,\bar{\undertilde{l}}_m)=E[\undertilde{Y}_k(\bar{\undertilde{a}}_m,\underline{\undertilde{0}})-\undertilde{Y}_k(\bar{\undertilde{a}}_{m-1},\underline{\undertilde{0}})|\bar{\undertilde{A}}_m=\bar{\undertilde{a}}_m,\bar{\undertilde{L}}_m=\bar{\undertilde{l}}_m].
\end{equation}
Here, $\undertilde{\gamma}^*_{m,k}(\bar{\textbf{a}}_m,\bar{\textbf{l}}_m)$ is a vector of effects on all of the outcomes in the cluster.  \citet{jamie1994} calls these contrasts `blip' functions because they capture the conditional effects of one last blip of (cluster level) exposure at time $m$ ($\undertilde{a}_m$) followed by no further direct or indirect exposure to any unit in the cluster thereafter compared to no direct or indirect exposure to any unit in the cluster at time $m$ or thereafter. We will denote by $\undertilde{\gamma}^*_{m,k,j}(\bar{\textbf{a}}_m,\bar{\textbf{l}}_m)$ the effect on cluster unit $j$.

To fix ideas, in the ophthalmological application considered by \citet{he2015structural}, $A_{i,m,j}$ indicated whether eye $j$ in subject $i$ received topical antihypertensive medication at time $m$. $Y_{i,k,j}$ was the intraocular pressure in eye $j$ of subject $i$ at time $k$. $\undertilde L_{i,m}$ comprised a vector of subject and eye-specific covariates measured at time $m$. $\undertilde\gamma(\undertilde{\bar{a}}_m,\undertilde{\bar{l}}_j)$ was the conditional effect of one last application of the antihypertensive at doses $\undertilde a_m = (a_{m,1},a_{m,2})$ at time $m$ on the vector of intraocular pressures at time $k>m$, given exposure and covariate history through time $m$.

\subsection{Identification}
Identification and estimation in the cluster interference setting will proceed almost identically to the no interference case, but with variables grouped at the cluster level. We first make the standard causal consistency assumption adapted for cluster interference:
\begin{assumption}\label{cluster_consistency}(Cluster Consistency) $\undertilde{\bar{Y}}_{i,\tau}(\bar{\undertilde{a}}_{i,\tau-1})=\undertilde{\bar{Y}}_{i,\tau}$ if $\bar{\undertilde{A}}_{i,\tau-1}=\bar{\undertilde{a}}_{i,\tau-1}$
\end{assumption}
This assumption states that the potential outcomes in a cluster under any given particular cluster-level exposure regime $\bar{\undertilde{a}}_{i,\tau-1}$ are equal to the observed outcomes if the observed exposure regime in that cluster was actually $\bar{\undertilde{a}}_{i,\tau-1}$. The consistency assumption links the potential outcomes to the observed data.

We next make the conditional parallel trends assumption at the cluster level:
\begin{assumption}\label{parallel_trends_cluster} (Cluster-level conditional parallel trends)
\[
E[\undertilde{Y}_k(\bar{\undertilde{a}}_{m-1},\underline{\undertilde{0}})-\undertilde{Y}_{k-1}(\bar{\undertilde{a}}_{m-1},\underline{\undertilde{0}})|\bar{\undertilde{A}}_m,\bar{\undertilde{L}}_m] = E[\undertilde{Y}_k(\bar{\undertilde{a}}_{m-1},\underline{\undertilde{0}})-\undertilde{Y}_{k-1}(\bar{\undertilde{a}}_{m-1},\underline{\undertilde{0}})|\bar{\undertilde{A}}_{m-1},\bar{\undertilde{L}}_m]
\] almost surely for all $k>m$ 
\end{assumption}
This assumption states that conditional on cluster covariate and exposure history, the cluster exposure assignment is mean independent of the vector of unexposed (i.e., no exposed units in the cluster from time $m$ onward) potential outcome trends. It would be violated if clusters receiving certain exposures at a given time tended to differ from clusters receiving no exposure at that time in ways that are associated with future untreated trends. In the eye example, perhaps subjects stopped taking the medication because they somehow knew they would stabilize in the absence of treatment (a flat trend), while those who continued to take medication did so because they somehow knew they would get worse in the absence of medication (a positive trend). 

Finally, we make a cluster version of the positivity assumption:
\begin{assumption}\label{cluster_positivity}
    (Cluster Positivity) $f_{\undertilde{A}_m|\undertilde{\bar{L}}_m,\undertilde{\bar{A}}_{m-1}}(\undertilde{0}|\bar{\undertilde{l}}_m,\bar{\undertilde{a}}_{m-1})>0$ whenever $f_{\bar{\undertilde{L}}_m,\bar{\undertilde{A}}_{m-1}}(\bar{\undertilde{l}}_m,\bar{\undertilde{a}}_{m-1})>0.$
\end{assumption}
This positivity assumption states that for any covariate and exposure history in a cluster, it is always possible that all units in the cluster will be unexposed at any given time.

To show that $\undertilde\gamma^*$ is identified under our assumption set, we begin by defining 
\begin{equation}
\undertilde{H}_{m,k}(\undertilde\gamma) = \undertilde{Y}_k-\sum_{j=m}^{k-1}\undertilde\gamma_{j,k}(\bar{\undertilde{A}}_j,\bar{\undertilde{L}}_j) \text{ for }  k>m \text{ and } \undertilde{H}_{t,t}\equiv \undertilde{Y}_t. 
\end{equation}
We refer to $\undertilde{H}_{m,k}(\undertilde\gamma)$ as `blipped down' outcomes. $\undertilde{H}_{m,k}(\undertilde\gamma)$ is a useful quantity because by arguments in \citet{jamie1994} and \citet{he2015structural}, under the cluster consistency assumption (\ref{cluster_consistency}), 
\begin{equation}\label{robins1994cluster}
E[\undertilde{H}_{m,k}(\undertilde{\gamma}^*)|\bar{\undertilde{A}}_m,\bar{\undertilde{L}}_m] = E[\undertilde{Y}_k(\bar{\undertilde{A}}_{m-1},\undertilde{\underline{0}})|\bar{\undertilde{A}}_m,\bar{\undertilde{L}}_m].
\end{equation}
That is, outcomes blipped down by the true blip function behave in conditional expectation like unexposed potential outcomes $\undertilde{Y}_k(\bar{\undertilde{A}}_{m-1},\undertilde{\underline{0}})$. Together with the parallel trends assumption \ref{parallel_trends_cluster}, (\ref{robins1994cluster}) implies that 
\begin{align}
\begin{split}\label{H_parallel_cluster}
E[\undertilde{H}_{m,k}(\undertilde\gamma)-\undertilde{H}_{m,k-1}(\undertilde\gamma)|\bar{\undertilde{A}}_m=\bar{\undertilde{a}}_m,\bar{\undertilde{L}}_m]=\\
E[\undertilde{H}_{m,k}(\undertilde\gamma)-\undertilde{H}_{m,k-1}(\undertilde\gamma)|\bar{\undertilde{A}}_m=(\bar{\undertilde{a}}_{m-1},\undertilde{0}),\bar{\undertilde{L}}_m].
\end{split}
\end{align}
For alternative choices of $\undertilde\gamma$, it can be tested whether (\ref{H_parallel_cluster}) with $\undertilde\gamma$ in place of the true $\undertilde\gamma^*$ holds in the observed data distribution. Thus, as in Theorem 3 of \citet{shahn2022structural}, (\ref{H_parallel_cluster}) is the basis for identification of $\undertilde\gamma$ and construction of consistent estimating equations. In particular, $\undertilde\gamma^*$ is the unique solution to

\begin{equation}  \label{np_est_eq_cluster}
E\left[ \sum_{\tau\geq k> m}\undertilde{U}_{m,k}(s_{m},\undertilde{\gamma})\right] =0
\end{equation}
where 
\begin{align*}
\undertilde{U}_{m,k}(s_{m},\undertilde{\gamma}) & =\{\undertilde{H}_{m,k}( 
\undertilde{\gamma})-\undertilde{H}_{m,k-1}(\undertilde{\gamma})\}\times
\\
& \{s_{m}(k,\bar{\undertilde{A}}_{m},\bar{\undertilde{L}}_m)-E[s_{m}(k,\bar{\undertilde{A}}_{m},\bar{\undertilde{L}}_m)|%
\bar{\undertilde{A}}_{m-1},\bar{\undertilde{L}}_m] \}
\end{align*} for all functions $s_{m}(k,\bar{\undertilde{A}}_{m},\bar{\undertilde{L}}_m)$ of $k$, $\bar{\undertilde{A}}_{m}$, and $\bar{\undertilde{L}}_m$. \newline
\newline
Further, $\undertilde\gamma^*$ satisfies:
\begin{equation}\label{estimating_equations_cluster}
E[\sum_{\tau\geq k> m}\sum_{m=0}^{\tau-1}\undertilde{U}^{\dagger}_{m,k}(s_{m},\undertilde{\gamma}^*,\widetilde{v}_{m},\widetilde{E}_{\undertilde{A}_m|\overline{\undertilde{L}}_m,\bar{\undertilde{A}}_{m-1}})] =0
\end{equation}
with 
\begin{align*}
&\undertilde{U}^{\dagger}_{m,k}(s_{m},\undertilde{\gamma}^{*}) =\{\undertilde{H}_{m,k}(\undertilde{\gamma}%
^{*})-\undertilde{H}_{m,k-1}(\undertilde{\gamma}^{*})-\widetilde{v}_{m}(k,\overline{\undertilde{L}}_{m},%
\overline{\undertilde{A}}_{m-1})\}\times \\
& \{s_{m}(k,\overline{\undertilde{L}}_{m},\overline{\undertilde{A}}_{m})-\widetilde{E}_{\undertilde{A}_{m}|%
\overline{\undertilde{L}}_{m},\overline{\undertilde{A}}_{m-1}}[ s_{m}(k,\overline{\undertilde{L}}_{m},\overline{\undertilde{A}}%
_{m})|\overline{\undertilde{L}}_{m},\overline{\undertilde{A}}_{m-1}] \}
\end{align*}
for any $s_{m}(k,\overline{\undertilde{l}}_{m},\bar{\undertilde{a}}_{m})$ if either

(a) $\widetilde{v}_{m}(k,\overline{\undertilde{l}}_{m},\bar{\undertilde{a}}_{m-1})=v_{m}(k,\overline{\undertilde{l}}_{m},\bar{\undertilde{a}}_{m-1};\undertilde\gamma^{*})$ 
for all $k>m$, where 
\begin{equation*}
v_{m}(k,\overline{\undertilde{l}}_{m},\bar{\undertilde{a}}_{m-1};\undertilde\gamma)\equiv E[\undertilde{H}_{m,k}(\undertilde\gamma) - \undertilde{H}_{m,k-1}(\undertilde\gamma)|\bar{\undertilde{L}}_m,\bar{\undertilde{A}}_{m-1}]
\end{equation*}
or

(b) $\widetilde{E}_{\undertilde{A}_{m}|\bar{\undertilde{L}}_{m},\bar{\undertilde{A}}_{m-1}}=E_{\undertilde{A}_{m}|\bar{\undertilde{L}}%
_{m},\bar{\undertilde{A}}_{m-1}}$ for all m.\newline
\newline
(\ref{estimating_equations_cluster}) and is thus a doubly robust estimating
function. 

It follows from (\ref{robins1994cluster}) that, given $\undertilde\gamma^*$, the expected outcome trajectory under no exposure for any unit, i.e. $E[\undertilde{Y}_k(\bar{\undertilde{0}})]$, is identified as $E[\undertilde{H}_{0,k}(\undertilde\gamma^*)]$ for each $k$. Other marginal quantities of interest are also identified as means of $\undertilde\gamma^*_{m,k}(\bar{\undertilde{A}}_m,\bar{\undertilde{L}}_m)$ in various subpopulations. For example, the mean of $\undertilde\gamma^*_{m,k,j}(\bar{\undertilde{A}}_m,\bar{\undertilde{L}}_m)$ among units with $A_{m,j}=1$ and $A_{m,j'}=0$ for all $j'\neq j$ identifies $E[\undertilde Y_{k,j}(\bar{\undertilde{A}}_{m-1},\undertilde{A}_m=(0,\ldots,0,A_{m,j},0,\ldots,0),\underline{\undertilde{0}})-\undertilde{Y}_{k,j}(\bar{\undertilde{A}}_{m-1},\underline{\undertilde{0}})]$, i.e., the conditional average effect of a last blip of direct exposure to unit $j$ (e.g., the left eye) at time $m$ on the outcome in unit $j$ at time $k$ among those only directly exposed at $m$, marginalizing over direct and indirect exposure and covariate history. This quantity can be computed separately for particular values of $j$ if meaningful (e.g., if clusters are households and direct effects may differ between parents and children), or averaged over values of $j$ to obtain a measure of average direct effect. Similarly, the mean of $\undertilde\gamma^*_{m,k,j}(\bar{\undertilde{A}}_m,\bar{\undertilde{L}}_m)$ among units with $A_{m,j}=0$ and $A_{m,j'}\neq 0$ for at least one $j'\neq j$ identifies $E[\undertilde Y_{k,j}(\bar{\undertilde{A}}_{m-1},\undertilde{A}_m=(A_{m,1},\ldots,A_{m,j-1},0,A_{m,j+1},\ldots,A_{m,J}),\underline{\undertilde{0}})-\undertilde Y_{k,j}(\bar{\undertilde{A}}_{m-1},\underline{\undertilde{0}})]$, i.e., the average effect of a last blip of indirect exposure to unit $j$ (e.g., the left eye) at time $m$ on the outcome in unit $j$ at time $k$ among those indirectly exposed at $m$, marginalizing over the nature of the indirect exposure at time $m$ (i.e., which other units were exposed and at which levels) and direct and indirect exposure history. Again, if certain values of $j$ are meaningfully different from others, this quantity can be computed separately for each $j$. Alternatively, the quantity can be averaged over values of $j$. Hopefully, it is clear that many potentially useful causal estimands are byproducts of the blip function.

\subsection{Estimation}
\subsubsection{Parametric Blip Model Specification}
When there are few members in each cluster, few time points, and few covariates, it might be possible to specify a saturated blip model capturing full interactions between direct and spillover effects. For example, suppose that clusters each have two units, exposures are binary and staggered, there are just two time exposure time points, and there are no covariates. Then one might specify the blip models 
\begin{align*}
    \undertilde\gamma_{0,k}(\undertilde{a}_0;\psi)=\begin{bmatrix}
           \psi_{0,k,1}^1 a_{10} + \psi_{0,k,1}^2 a_{20} + \psi_{0,k,1}^3 a_{10}a_{20} \\
           \psi_{0,k,2}^1 a_{10} + \psi_{0,k,2}^2 a_{20} + \psi_{0,k,2}^3 a_{10}a_{20}
         \end{bmatrix}
\end{align*}
\begin{align*}
    \undertilde\gamma_{1,2}(\bar{\undertilde{a}}_1;\psi)=\begin{bmatrix}
           \psi_{1,2,1}^1 a_{11} + \psi_{1,2,1}^2 a_{12} + \psi_{1,2,1}^3 a_{11}a_{12} +\psi_{1,2,1}^4 a_{11}a_{02} + \psi_{1,2,1}^5 a_{12}a_{01} \\
           \psi_{1,2,2}^1 a_{11} + \psi_{1,2,2}^2 a_{12} + \psi_{1,2,2}^3 a_{11}a_{12} +\psi_{1,2,2}^4 a_{11}a_{02} + \psi_{1,2,2}^5 a_{12}a_{01} 
         \end{bmatrix}
\end{align*}
where $a_{m,j}$ is the counterfactual exposure at time $m$ for unit $j$ within each cluster, and $\psi_{m,k,j}^h$ is the $h^{th}$ coefficient of the blip function determining the effects of exposures applied at time $m$ on the outcome at time $k$ (taking values 1 or 2) in unit $j$ of the cluster. Under these models, the effect of a blip of exposure applied to either unit in the cluster on the outcome of either unit in the cluster can depend on: the time at which exposure was applied, the current exposure and exposure history of the other unit, the time at which the outcome is to be assessed, and the index of the unit in the cluster (e.g., left or right eye) on which the effect is to be assessed. 

When there are many units in a cluster, many time points, or high dimensional or continuous covariates or exposures, simplifying assumptions must be encoded in the blip model. For example, perhaps the model for the effect of a cluster exposure assignment on any unit can be assumed to depend only on the unit's direct exposure and some summary (e.g., average) of the exposures of the other units in the cluster. The considerations around specification of these simplifying models are similar to those regarding specification of the exposure mapping under network interference structures, which we return to later.

\begin{remark}
As we mentioned in the introduction, many applications of time-varying DiD methods have been in the context of `staggered adoption' exposures. In these settings, it is frequently the case that initial direct exposure is an absorbing state. That is, once direct exposure is initiated, its level is maintained throughout follow up. (In the case of an absorbing binary exposure, like Medicaid expansion, exposure can only be started once, and it is never stopped once it is started.) Formally, one natural way to code an absorbing exposure state is $A_{m,j}=A_{t,j}$ if $A_{t,j}\neq 0$ and $m>t$. Under this coding, the positivity assumption (Assumption \ref{cluster_positivity}), which requires positive probability of 0 exposure under all possible histories, fails. This is because if $\sum \bar{a}_{m-1,j}>0$, then exposure $A_{m,j}$ cannot be equal to 0, implying $\undertilde A_m$ cannot be equal to 0. 

However, when initial exposure is absorbing, we can recode the direct exposure trajectory of each unit such that it takes a non-zero value only at the time of initiation.  For example, if a unit's direct exposure history were $(0,0,1,1,1)$ (with the latter two 1's deterministically following from the first 1), it could be coded as $(0,0,1,0,0)$ without any loss of information. This recoding restores the possibility of positivity. The recoding is also useful even if exposure switches on and off in the data but interest centers only on the effect of initiation. See \citet{shahn2022structural} for more discussion and an application.
\end{remark}

\begin{remark}
    When exposures are continuous, not usually equal to 0, and jump around over time (for example, average temperatures), recoding can also be helpful. In particular, exposure at time $m$ can be coded as the change in absolute level of exposure from time $m-1$, with the absolute level of the exposure at $m-1$ included as part of the covariate $\bar{L}_{m,j}$. Then $\undertilde\gamma_{m,k}( \bar{\undertilde{a}}_m,\bar{\undertilde{l}}_m)$ represents the conditional lasting effect of changes in exposure by $\undertilde a_m$, at time $m$ followed by no further changes, conditional on the history of absolute exposure levels prior to $m$. See \citet{shahn2022structural} for more discussion and an application.
\end{remark}

\subsubsection{Estimating equations and standard errors} \label{cluster_se}
With the blip models $\undertilde\gamma_{m,k}(\bar{\undertilde{a}}_m,\bar{\undertilde{l}}_m;\psi)$ correctly specified, consistent estimation can be achieved by solving the estimating equations (\ref{estimating_equations_cluster}) with consistent multivariate regression estimators $\hat{E}_{\undertilde{A}_m|\bar{\undertilde{L}}_m,\bar{\undertilde{A}}_{m-1}}$ and $\hat{v}_m(k,\bar{\undertilde{l}}_m,\bar{\undertilde{a}}_{m-1};\gamma(\psi))$ of the nuisance functions $E_{\undertilde{A}_m|\bar{\undertilde{L}}_m,\bar{\undertilde{A}}_{m-1}}$ and $v_m(k,\bar{\undertilde{l}}_m,\bar{\undertilde{a}}_{m-1};\gamma(\psi))$ plugged in.  

From standard M-estimation theory (\citep{van2000asymptotic,stefanski2002calculus,ross2024m}) with clusters as (independent) units\citep{yuan1998asymptotics}, standard errors can be consistently estimated via non-parametric bootstrap (with clusters as the sampling units) or the empirical sandwich variance estimator. A consistent sandwich estimator of the sampling variance assuming nonparametric estimation of the nuisance models is given by  
\begin{equation}
 \widehat{\mathrm{Var}}(\hat\psi)
=P_n\{\hat{S}_{\psi,i}\hat{S}_{\psi,i}^T\}
\end{equation}
with 
$\widehat S_{\psi,i}=\widehat A^{-1} g_i(\hat\psi)$, $\widehat A=\left.\partial \bar g_n(\psi)/\partial\psi^\top\right|_{\psi=\hat\psi}$, and
$g_i(\psi)=\sum_{\tau\ge k>m}\sum_{m=0}^{\tau-1} U^\dagger_{m,k,i}(s_m,\gamma(\psi),\hat v_m,\widehat E_{A_m\mid L_m,\bar A_{m-1}})$ with $\textbf{U}^{\dagger}_{m,k}$ defined as in (\ref{estimating_equations_cluster}).


\section{Network Interference}\label{network}
\subsection{Notation and Effects of Interest}
Suppose we observe a cohort of $n$ units at time points $1,\ldots,\tau$. Let $A_{i,k}$ denote the exposure assignment of unit $i$ at time $k$ taking values in $\mathcal{A}$. Let $\textbf{A}_k$ denote the full cohort exposure assignment for all units at time $k$. Let $L_{i,k}$ denote covariates and $Y_{i,k}$ denote the outcome of unit $i$ at time $k$. Let $Y_{i,k}(\bar{\textbf{a}}_k)$ denote the potential outcome of unit $i$ at time $k$ under the full cohort exposure regime $\bar{\textbf{a}}_k$. Following \citet{aronow2017estimating} and \citet{manski2013identification}, we will assume that $Y_{i,k}(\bar{\textbf{a}}_k)$ depends only on some `exposure mapping' $\phi(\bar{\textbf{a}}_k,\theta_{i,k})$ of $\bar{\textbf{a}}_k$ and unit/time-specific characteristics $\theta_{i,k}$, i.e., we assume $Y_{i,k}(\bar{\textbf{a}}_k) = Y_{i,k}(\phi(\bar{\textbf{a}}_k,\theta_{i,k}))$. 

For tractability, we will make some further simplifying assumptions about the structure of $\phi(\bar{\textbf{a}}_k,\theta_{i,k})$. In particular, we assume that $\phi(\bar{\textbf{a}}_k,\theta_i)=(\phi_1(\textbf{a}_1,\theta_{i,1}),\ldots,\phi_k(\textbf{a}_k,\theta_{i,k}))$ with $\phi_j(\textbf{a}_j,\theta_{i,j})=(a_{i,j},h_{j}(\textbf{a}_j;\theta_{i,j}))$ and $h_{j} : \mathcal{A}^n\times \mathcal{\theta}\rightarrow \mathbf{R^p}$. That is, the exposure mapping $\phi(\bar{\textbf{a}}_k,\theta_{i,k})$ comprises separate mappings for each time point, and at each time point the total exposure $\textbf{a}_j$ only affects unit $i$ outcomes through the direct exposure $a_{i,j}$ and through a p-dimensional summary $h_{j}(\textbf{a}_j;\theta_{i,j})$ of the full cohort wide exposure. For example, $\theta_{i,k}$ might represent the indices of the neighboring units to unit $i$ at time $k$, and $h_{k}(\textbf{a}_j;\theta_{i,k})$ may simply denote the sum of the exposures in units neighboring unit $i$. Such an exposure mapping would encode the assumptions that interference is limited to neighboring units and that neighboring units have exchangeable additive interference effects. The implicit assumptions of this mapping could be violated if, for example, units are counties and exposures in neighboring counties with higher populations have larger spillover effects. In that case, a population-weighted sum of exposures in neighboring counties may be more appropriate.

In the following, to simplify notation, we let $D_{i,k}=\phi_k(\textbf{A}_k,\theta_{i,k})$ denote the total direct and indirect exposure received by unit $i$ at time $k$. We are interested in estimating the causal contrasts:
\begin{equation}\label{blip}
\gamma_{m,k}^*(\bar{d}_m,\bar{l}_m)=E[Y_k(\bar{d}_{m},\textbf{\underline{0}}))-Y_k(\bar{d}_{m-1},\textbf{\underline{0}}))|\bar{D}_m=\bar{d}_{m},\bar{L}_m=\bar{l}_m].
\end{equation}
 These are so-called 'blip' effects of receiving (combined direct and indirect) exposure $d_m$ at time $m$ followed by no (direct or indirect) exposure thereafter versus receiving no (direct or indirect) exposure at time $m$ or thereafter in units with (direct and indirect) exposure and covariate history $(\bar{d}_m,\bar{l}_m)$. The blip functions $\gamma_{m,k}^*$ are the contrasts modeled by a SNMM. (Note that while we say `no exposure', we mean exposure taking whatever value $0$ corresponds to in the exposure mapping. It is always possible to recode such that 0 refers to any exposure value of interest, as mentioned in \citet{jamie1994,jamie2004}.) 

\subsection{Identification}
First, we adapt the standard consistency assumption to accommodate interference via the mapped exposure.
\begin{assumption} \label{consistency}(Consistency) $Y_k(\overline{d}_k) = Y_k$ if $\overline{D}_k=\overline{d}_k$
\end{assumption}

Assumption \ref{consistency} states that a unit's potential outcome under a particular exposure mapping $\overline d_k$ equals its observed outcome if its observed (direct and indirect) exposure $\overline D_k=\overline d_k$, and would be violated if interference occurs in ways not captured by the exposure mapping. 

We further make the conditional parallel trends assumption:
\begin{assumption}\label{parallel_trends} (Network conditional parallel trends)
\begin{align*}
    E[&Y_k(\bar{d}_{m-1},\underline{0}))-Y_{k-1}(\bar{d}_{m-1},\underline{0}))|\bar{D}_m=\bar{d}_m,\bar{L}_m]\\
&=E[Y_k(\bar{d}_{m-1},\textbf{\underline{0}})-Y_{k-1}(\bar{d}_{m-1},\textbf{\underline{0}})|\bar{D}_m=(\bar{d}_{m-1},0),\bar{L}_m]
\end{align*}
for all $\bar{d}_m$, $m$, and $k$.
\end{assumption}
Assumption \ref{parallel_trends} states that average trends in potential outcomes under the regime $\overline D_m=(\bar d_{m-1}, \underline 0)$ are equal among units with the same covariate history through $m$ and exposure history $\bar d_{m-1}$, whether their exposure at time $m$ is equal to $d_m$ or $0$. That is, current exposure is conditionally mean independent of future unexposed potential outcome trends given covariate and exposure history. \ref{parallel_trends} is similar to parallel trends assumptions adopted in prior work in time-varying DiD settings (e.g., \citet{renson2023identifying, callaway2021difference}), and is identical to that adopted by \citet{shahn2022structural} except that it indexes potential outcomes by the exposure mapping rather than the unit's (direct) exposure only. As in \citet{shahn2022structural}, the parallel trends assumption can apply to a potentially multidimensional or continuous exposure.

Finally, we make a positivity assumption adapted to the exposure mapping setting: 
\begin{assumption}\label{positivity}
(Network positivity) $f_{D_m|\bar{L}_m,\bar{D}_{m-1}}(0|\bar{l}_m,\bar{d}_{m-1})>0 \text{ whenever } f_{\bar{L}_m,\bar{D}_{m-1}}(\bar{l}_m,\bar{d}_{m-1})>0.$
\end{assumption}
Assumption \ref{positivity} would be violated if there are no untreated (both directly and indirectly) units at time $k$ at a given level of prior covariate and exposure mapping levels. An example of a setting where Assumption \ref{positivity} would be violated is if every state with a Democratic governor implemented a policy at the same time, and having a Democratic governor was a covariate required to satisfy Assumption \ref{parallel_trends}.

To show identification of the true $\gamma^*$ from (\ref{blip}) under our assumption set, we begin by defining
\begin{equation}
H_{m,k}(\gamma) = Y_k-\sum_{j=m}^{k-1}\gamma_{j,k}(\bar{D}_j,\bar{L}_j) \text{ for } k>m \text{ and } H_{t,t}\equiv Y_t. 
\end{equation}
We refer to $H_{m,k}(\gamma)$ as the ``blipped down" outcome. This quantity is useful because $H_{m,k}(\gamma^*)$, i.e., the outcome blipped down by the \textit{true} blip function $\gamma^*$, has special properties. Specifically, under the Consistency Assumption \ref{consistency} alone, $H_{m,k}(\gamma^*)$ behaves in conditional expectation like the potential outcome $Y_k(\overline D_{m-1},\underline 0)$, i.e.,
\begin{equation}\label{robins1994}
E[H_{m,k}(\gamma)|\bar{D}_m,\bar{L}_m] = E[Y_k(\bar{D}_{m-1},\textbf{\underline{0}})|\bar{D}_m,\bar{L}_m].
\end{equation}
Together with Assumption \ref{parallel_trends}, (\ref{robins1994}) implies that 
\begin{align}
\begin{split}\label{H_parallel}
E[H_{m,k}(\gamma^*)-H_{m,k-1}(\gamma^*)|\bar{D}_m=\bar{d}_m,\bar{L}_m]=\\
E[H_{m,k}(\gamma^*)-H_{m,k-1}(\gamma^*)|\bar{D}_m=(\bar{d}_{m-1},\textbf{0}),\bar{L}_m].
\end{split}
\end{align}
For alternative choices of $\gamma$, it can be tested whether (\ref{H_parallel}) with $\gamma$ in place of the true $\gamma^*$ holds in the observed data distribution. Thus, as in Theorem 3 of \citet{shahn2022structural}, (\ref{H_parallel}) is the basis for identification of $\gamma^*$ and construction of consistent estimating equations. In particular, under Assumptions \ref{consistency}-\ref{parallel_trends}, $\gamma^*$ is the unique solution to

\begin{equation}  \label{np_est_eq}
E\left[ \sum_{\tau\geq k> m}U_{m,k}(s_{m},\mathbf{\gamma})\right] =0
\end{equation}
where 
\begin{align*}
U_{m,k}(s_{m},\mathbf{\gamma}) & =\{H_{m,k}( 
\mathbf{\gamma})-H_{m,k-1}(\mathbf{\gamma})\}\times
\\
& \{s_{m}(k,\bar{D}_{m},\bar{L}_m)-E[s_{m}(k,\bar{D}_{m},\bar{L}_m)|%
\bar{D}_{m-1},\bar{L}_m] \}
\end{align*}
as $s_{m}(k,\bar{D}_{m},\bar{L}_m)$ varies over all functions of $k$, $\bar{D}_{m}$, and $\bar{L}_m$. \newline
\newline
Further, $\gamma^*$ satisfies the estimating function 
\begin{equation}\label{estimating_equations}
E[\sum_{\tau\geq k> m}\sum_{m=0}^{\tau-1}U^{\dagger}_{m,k}(s_{m},\mathbf{\gamma}^*,\widetilde{v}_{m},\widetilde{E}_{D_m|\overline{L}_m,\bar{D}_{m-1}})] =0
\end{equation}
with 
\begin{align*}
&U^{\dagger}_{m,k}(s_{m},\mathbf{\gamma}^{*}) =\{H_{m,k}(\mathbf{\gamma}%
^{*})-H_{m,k-1}(\mathbf{\gamma}^{*})-\widetilde{v}_{m}(k,\overline{L}_{m},%
\overline{D}_{m-1})\}\times \\
& \{s_{m}(k,\overline{L}_{m},\overline{D}_{m})-\widetilde{E}_{A_{m}|%
\overline{L}_{m},\overline{D}_{m-1}}[ s_{m}(k,\overline{L}_{m},\overline{D}%
_{m})|\overline{L}_{m},\overline{D}_{m-1}] \}
\end{align*}
for any $s_{m}(k,\overline{l}_{m},\bar{d}_{m})$ if either

(a) $\widetilde{v}_{m}(k,\overline{l}_{m},\bar{d}_{m-1})=v_{m}(k,\overline{l}_{m},\bar{d}_{m-1};\gamma^{*})$ 
for all $k>m$, where 
\begin{equation*}
v_{m}(k,\overline{l}_{m},\bar{d}_{m-1};\gamma)\equiv E[H_{m,k}(\gamma) - H_{m,k-1}(\gamma)|\bar{L}_m,\bar{D}_{m-1}]
\end{equation*}
or

(b) $\widetilde{E}_{D_{m}|\overline{L}_{m},\bar{D}_{m-1}}=E_{D_{m}|\overline{L}%
_{m},\bar{D}_{m-1}}$ for all m.\newline
\newline
(\ref{estimating_equations}) is thus a doubly robust estimating function. 

It follows from (\ref{robins1994}) that, given $\gamma^*$, the expected outcome trajectory under no exposure for any unit, i.e., $E[Y_k(\bar{\textbf{0}})]$, is identified as $E[H_{0,k}(\gamma^*)]$ for each $k$. Other marginal quantities of interest are also identified as expectations of $\gamma^*_{m,k}(\bar{a}_m,\bar{h}_m)$ in various subpopulations. For example, the mean of $\gamma_{m,k}^*(\bar{D}_m,\bar{L}_m)$ among units with $h_m=1$ and $a_m=0$ identifies $E[Y_{k}(\bar{D}_{m-1},{d}_m=(0,h_m),\underbar{0})-Y_{k}(\bar{D}_{m-1},\underbar{0})]$, i.e., the average effect of a last blip of indirect exposure at $m$ on the outcome at $k$ among those indirectly exposed and not directly exposed at $m$, marginalizing over direct and indirect exposure history. Similarly, the mean of $\gamma_{m,k}^*(\bar{D}_m,\bar{L}_m)$ among units with $a_m=1$ and $h_m=0$ identifies $E[Y_{k}(\bar{D}_{m-1},{d}_m=(a_m,0),\underbar{0})-Y_{k}(\bar{D}_{m-1},\underbar{0})]$, i.e., the average effect of a blip of direct exposure at $m$ on the outcome at $k$ among those directly exposed and not indirectly exposed at time $m$, marginalizing over direct and indirect exposure and covariate history. In the presence of covariates, $\gamma^*$ also characterizes how all of the above quantities vary with time varying covariates. So, a myriad of causal estimands are identified as byproducts of the blip function.

\subsection{Estimation}
\subsubsection{Parametric Blip Model Specification}
In practice, estimation entails specification of some finite-dimensional (though possibly saturated, as in our simulations) parametric blip models 
\begin{equation}
\gamma_{m,k}(\bar{d}_m,\bar{l}_m)=E[Y_k(\bar{d}_{m},\textbf{\underline{0}})-Y_k(\bar{d}_{m-1},\textbf{\underline{0}})|\bar{D}_m=\bar{d}_{m},\bar{L}_m=\bar{l}_m] = \gamma_{m,k}(\bar{d}_m,\bar{l}_m;\psi)
\end{equation}
with parameter $\psi$ such that $\gamma_{m,k}(\bar{d}_m,\bar{l}_m;\psi)=0$ whenever $d_m=0$ or $\psi=0$. For example, suppose direct exposure $A_m$ is binary and: $d_m=(0,0)$ if $A_m=0$ and no neighboring units are directly treated; $d_m=(0,1)$ if $A_m=0$ and at least one neighboring unit is directly treated; $d_m=(1,0)$ if $A_m=1$ and no neighboring units are directly treated; and $d_m=(1,1)$ if $A_m=1$ and at least one neighboring unit is directly treated. One possible blip model specification could be 
\begin{equation}
    \gamma_{m,k}(\bar{d}_m,\bar{l}_m;\psi) = \psi_1a_m + \psi_2h_{m} + \psi_3a_mh_{m} +\psi_4a_ml_m
\end{equation}
Under this blip model specification, the effect of a blip of direct and/or indirect exposure does not depend on exposure history. However, the effect of a blip of direct exposure depends both on whether indirect exposure occurs at the same time (via $\psi_3$) and on covariates at the time of exposure (via $\psi_4$). The effect of a blip of indirect exposure varies depending on whether direct exposure occurs at the same time (via $\psi_3$). 

Alternatively, suppose we wish to encode dependence on past exposure history. We could specify, for example, the blip function

\begin{equation}\label{example3}
    \gamma_{m,k}(\bar{d}_m,\bar{l}_m;\psi) = \psi_1 a_m + \psi_2 h_{m} + \psi_3a_mh_{m} +\psi_4a_ml_m + \psi_5(\sum_{j=1}^{m-1} a_j)h_m
\end{equation}
in which the effect of (present) indirect exposure depends on the sum of past direct exposure.

In the example blip model specifications above, the parameter $\psi$ is shared across blip models for effects of exposures delivered at different times. Let $\psi_{m,k}$ and $\psi_{m',k}$ denote the parameters of $\gamma_{m,k}$ and $\gamma_{m',k}$, respectively. We can also of course specify blip models such that parameters $\psi_{m,k}$ and $\psi_{m',k}$ are variation independent, e.g.,
\begin{equation}
    \gamma_{m,k}(\bar{d}_m,\bar{l}_m;\psi) = \psi_1^{m,k} a_m + \psi_2^{m,k} h_{m} + \psi_3^{m,k}a_mh_{m} +\psi_4^{m,k}a_ml_m + \psi_5^{m,k}(\sum_{j=1}^{m-1} a_j)h_m.
\end{equation}

Note that Remarks 1 and 2 from the cluster interference section (regarding recoding exposure variables to accommodate staggered adoption or continuous exposure settings) also apply here.

\subsubsection{Estimating equations and standard errors}\label{sec:network_esteq_se}
With the blip models $\gamma_{m,k}(\bar{d}_m,\bar{l}_m;\psi)$ correctly specified, consistent estimation can be achieved by solving the estimating equations (\ref{estimating_equations}) with consistent estimators $\hat{E}_{D_m|\bar{L}_m,\bar{D}_{m-1}}$ and $\hat{v}_m(k,\bar{l}_m,\bar{d}_{m-1};\gamma(\psi))$ of the nuisance functions $E_{D_m|\bar{L}_m,\bar{D}_{m-1}}$ and $v_m(k,\bar{l}_m,\bar{d}_{m-1};\gamma(\psi))$ plugged in. 
Our proposed estimator is an M-estimator. Standard theory for M-estimators \citep{van2000asymptotic,stefanski2002calculus,ross2024m} assumes that observations are independent. However, under interference, this assumption no longer holds. Therefore, standard sandwich and bootstrap estimators of the standard error are not necessarily consistent. 

Recent work in this area provides a way forward. Under regularity conditions for dependency-graph CLTs
\citep{kojevnikov2021limit,jetsupphasuk2025estimating},
the estimator admits the linearization
\[
\sqrt{n}(\hat\psi - \psi_0)
\;\rightsquigarrow\; \mathcal{N}\!\big(0,\; G^{-1}\Sigma G^{-\top}\big),
\]
where
\[
G \;=\; E\!\left[ \frac{\partial S_i(\psi_0)}{\partial \psi^\top} \right],
\qquad
\Sigma \;=\; \sum_{s=0}^\infty \Omega(s),
\]
with
\[
\Omega(s) \;=\; \lim_{n\to\infty}\frac{1}{n}
\sum_{i=1}^n\sum_{j\in\partial_n(i;s)}
E\!\big[\,S_i(\psi_0)S_j(\psi_0)^\top\big],
\]
\[
S_i=\sum_{\tau\geq k> m}\sum_{m=0}^{\tau-1}U^{\dagger}_{i,m,k}(s_{m},\mathbf{\gamma}^*,\widetilde{v}_{m},\widetilde{E}_{D_m|\overline{L}_m,\bar{D}_{m-1}}),
\]
and $\partial_n(i;s)$ denotes the set of units at graph distance $s$ from $i$. A consistent plug-in estimator can be constructed using a network-HAC approach:
\[
\hat V(\hat\psi) \;=\; \frac{1}{n}\,\hat G^{-1}\,\hat\Sigma\,\hat G^{-\top},
\]
where
\[
\hat G \;=\; \frac{1}{n}\sum_{i=1}^n
\left.\frac{\partial S_i(\psi)}{\partial \psi^\top}\right|_{\psi=\hat\psi},
\quad
\hat\Sigma \;=\; \sum_{s=0}^{S_n} \kappa\!\left(\tfrac{s}{b_n}\right)\hat\Omega_n(s),
\]
\[
\hat\Omega_n(0) \;=\; \frac{1}{n}\sum_{i=1}^n \hat S_i \hat S_i^\top,
\qquad
\hat\Omega_n(s) \;=\; \frac{1}{n}\sum_{i=1}^n
\sum_{j\in\partial_n(i;s)} \hat S_i \hat S_j^\top,\;\; s\ge 1,
\]
for $\hat S_i = S_i(\hat\psi)$, kernel $\kappa$ (e.g.\ Bartlett), and bandwidth $b_n\to\infty$ with $b_n/\sqrt{n}\to 0$. Standard errors are then obtained as
$\mathrm{SE}(\hat\psi_j) = \sqrt{\hat {V}_{j,j}(\hat\psi)}$ for $\hat{V}_{j,j}$ the $j^{th}$ element of $diag(\hat{V})$.

Another alternative to estimate uncertainty under network interference is the moving block bootstrap (MBB) \citep{kunsch1989jackknife,liu1992moving,ogburn2024causal}. Let units be ordered along their natural indexing (e.g., a line network, as in our simulation in the following section). Fix a block length $L$ and form circular, overlapping blocks of consecutive indices of length $L$. For each bootstrap replicate, draw $B=\lceil N/L\rceil$ block start positions independently and concatenate the corresponding blocks (with wrap–around) until $N$ observations are obtained. On each resample we rebuild the estimating dataset, re–fit the nuisance components entering $E[\phi]$, and re–solve the same estimating equations to obtain $\hat\psi^{*}$. Standard errors are taken as the empirical standard deviations of $\{\hat\psi^{*(b)}\}_{b=1}^B$. The block length $L$ should be chosen to cover the dependence radius implied by the interference mechanism (e.g., if outcomes depend on neighbors up to graph distance two, take $L\ge 5$ in a line network).

\section{Simulations}\label{simulations}
R code for all simulations in this section is available at https://github.com/zshahn/did-snmm-interference.
\subsection{Network Interference}\label{subsection:network_sim}
We illustrate our approach in the network interference setting with a simulation of a two time-step data generating process. We conceive of interference according to a geographic proximity network where units are arranged in a straight line such that each unit's outcomes are impacted by both its own direct exposure assignment and the exposure assignments of its two neighbors on either side. (The edge units have spillover effects from only one neighbor.) We let the unit index $i$ indicate order on the line and define the exposure mapping $D_{i,m}=\phi_m(\textbf{A}_m,\theta_{i,m})=(A_{i,m},h_m=max(A_{i-1,m},A_{i+1,m}))$. That is, all that matters about neighboring units' exposures is whether \textit{any} neighboring unit is treated. If both neighboring units are treated, the spillover effect is the same as if just one were treated. We let exposure be absorbing and thus coded it as 1 only the first time it occurs as in Remark 1. 

The data generating process was as follows:

\begin{align}\label{dgm}
    \begin{split}
    &U_i\sim Bernoulli(0.5)\\
    &Y_{i,0}, Y_{i,1}(D_{i,0}=(0,0)), Y_{i,2}(\bar{D}_{i,1}=(\bar{0},\bar{0}))\sim N(U_i,0.1)\\
    &A_{i,0} \sim Bernoulli(0.3 + 0.2U_i)\\
    &A_{i,1} \sim (1-A_{i,0})Bernoulli(0.3+0.2U_i)\\
    &Y_{i,1}\sim N(Y_{i1}(0) + \gamma_{01}(A_{i,0},max(A_{i-1,0},A_{i+1,0}))\\
    &Y_{i,2} \sim N(Y_{i,2}(0) + \gamma_{0,2}(A_{i,0},max(A_{i-1,0},A_{i+1,0})) + \gamma_{1,2}(A_{i,1},max(A_{i-1,1},A_{i+1,1})),0.1)
    \end{split}
\end{align}

$U_i$ is an unobserved baseline confounder that makes a constant additive contribution to the expectations of the untreated potential outcomes, ensuring that parallel trends (\ref{parallel_trends}) holds. We specified saturated blip models:
\begin{align}
     &\gamma_{0,k}(a_0,h_1) = \psi_1a_0 + \psi_2h_0 + \psi_3a_0(k-1) + \psi_4h_0(k-1) + \psi_5a_0h_0 +\psi_6a_0h_0(k-1)\\
     &\gamma_{1,2}(\bar{a}_1,\bar{h}_1) = \psi_7a_1 + \psi_8h_1 + \psi_9a_1h_1 +  
     \psi_{1,0}a_1h_0 +\psi_{1,1}h_1a_0+\psi_{1,2}h_1h_0+\psi_{1,3}a_1h_1h_0.
\end{align}
The true value of $\psi$ was $(\psi_1=1,\psi_2=.5,\psi_3=-.1,\psi_4=-.1,\psi_5=-.2,\psi_6=-.05,\psi_7=1,\psi_8=.5,\psi_9=-.1,\psi_{1,0}=-.1,\psi_{1,1}=-.1,\psi_{1,2}=-.05,\psi_{1,3}=-.05$). Effects of direct and spillover exposures vary with time of exposure (variation between $\gamma_{1,1}$ and $\gamma_{2,2}$), time since exposure (driven by $\psi_3$, $\psi_4$, and $\psi_6$), and direct and indirect exposure history (driven by $\psi_{1,0}$, $\psi_{1,1}$, and $\psi_{1,2}$). Concurrent direct and indirect exposures also interact ($\psi_{5}$,$\psi_6$, and $\psi_9$).

We simulated 500 data sets of $N=5,000$ units according to (\ref{dgm}). We then proceeded to estimate $\psi$ nonparametrically by solving estimating equations (\ref{estimating_equations}) with plugged in saturated sample average estimators $\hat{E}_{D_m|\bar{D}_{m-1}}$ and $\hat{v}_m(k,\bar{d}_{m-1};\gamma(\psi))$ of the nuisance functions $E_{D_m|\bar{D}_{m-1}}$ and $v_m(k,\bar{d}_{m-1};\gamma(\psi))$. We estimated standard errors using the moving block bootstrap procedure described in Section \ref{sec:network_esteq_se} with $L=5$ and 500 bootstrap replicates. All parameters and derived effects of interest were approximately unbiasedly estimated, and confidence interval coverage was near nominal. See Table \ref{table1}.  

\begin{table}[t!]
\centering
\caption{Simulation results for estimates of direct and spillover effects in a simple geographic proximity type network interference structure. 500 simulations at $N=5{,}000$, bootstrap blocks of size 5.}
\begin{tabular}{lccc}
\hline
\textbf{Estimand} & \textbf{True} & \textbf{Mean (sd)} & \textbf{95\% perc. cov.} \\
\hline
$\gamma_{0,1}(a_0=1,h_0=0)$                     & 1.00 & 1.00 (.006) & 0.942 \\
$\gamma_{0,1}(a_0=1,h_0=1)$                     & 1.30 & 1.30 (.005) & 0.942 \\
$\gamma_{0,1}(a_0=0,h_0=1)$                     & 0.50 & 0.50 (.005) & 0.952 \\
$\gamma_{0,2}(a_0=1,h_0=0)$                     & 0.90 & 0.90 (.009) & 0.944 \\
$\gamma_{0,2}(a_0=1,h_0=1)$                     & 1.05 & 1.05 (.009) & 0.946 \\
$\gamma_{0,2}(a_0=0,h_0=1)$                     & 0.40 & 0.40 (.008) & 0.926 \\
$\gamma_{1,2}(\bar a_1=(0,1),\,h_1=(0,0))$      & 1.00 & 1.00 (.014) & 0.938 \\
$\gamma_{1,2}(\bar a_1=(0,1),\,h_1=(1,0))$      & 0.90 & 0.90 (.008) & 0.930 \\
$\gamma_{1,2}(\bar a_1=(0,1),\,h_1=(0,1))$      & 1.40 & 1.40 (.012) & 0.920 \\
$\gamma_{1,2}(\bar a_1=(0,1),\,h_1=(1,1))$      & 1.20 & 1.20 (.011) & 0.944 \\
$\gamma_{1,2}(\bar a_1=(0,0),\,h_1=(0,1))$      & 0.45 & 0.45 (.008) & 0.940 \\
$\gamma_{1,2}(\bar a_1=(0,0),\,h_1=(1,1))$      & 0.40 & 0.40 (.009) & 0.930 \\
$\gamma_{1,2}(\bar a_1=(1,0),\,h_1=(0,1))$      & 0.35 & 0.35 (.008) & 0.916 \\
\hline
\end{tabular}\label{table1}
\end{table}

For illustrative purposes, we describe many of the effects from Table 1 in words. The immediate effect on the treated of direct but no indirect exposure at time 0 versus no exposure on the outcome at time 1 (i.e., $\gamma_{0,1}(d_0=(1,0))$) is equal to 1. The immediate effect on the treated of both direct and indirect exposure at time 0 versus no exposure on the outcome at time 1 (i.e., $\gamma_{0,1}(d_0=(1,1))$) is equal to 1.3. The immediate effect on the treated of indirect but no direct exposure at time 0 versus no exposure on the outcome at time 1 (i.e., $\gamma_{0,1}(d_0=(0,1))$) is equal to 0.5. Thus, at time 0, immediate effects of both direct and indirect exposure were positive, with spillover effects greater in the absence of direct exposure. 

Now we will go through the effects of blips of exposure at time 0 on outcomes at time 2. The effect on the treated of direct exposure at time 0 and no indirect exposure ever versus no exposure ever on the outcome at time 2 (i.e., $\gamma_{0,2}(d_0=(1,0))$) is equal to 0.8. The effect on the treated of direct and indirect exposure at time 0 and no exposure at time 1 versus no exposure ever on the outcome at time 2 (i.e., $\gamma_{0,2}(d_0=(1,1))$) is equal to 1. The effect on the treated of indirect exposure at time 0 and no direct exposure at times 0 or 1 versus no exposure ever on the outcome at time 2 (i.e., $\gamma_{0,2}(d_0=(0,1))$) is equal to 0.3. Thus, both direct and spillover effects wane over time compared to immediate effects. Further, the long term effects of blips of direct and indirect exposure are both positive, but the average effect of a blip of combined direct and indirect exposure was less than the sum of the effects of direct and indirect exposure alone. However, absent effect homogeneity assumptions, comparisons between blip functions for different exposures at time $m$ are not causal contrasts since these are effects on the treated and only apply to the distinct populations that followed those regimes at time $m$.

We note that given a consistent estimator $\hat{\psi}$, we are able to compute corresponding consistent estimators of other derived quantities of interest. An important example is that we can compute the consistent estimator $\hat{E}[\bar{Y}(\bar{0},\bar{0})]$ of the expected outcome trajectory had no unit received direct or indirect exposure as the sample average $\mathbf{P}_n H_{0,k}(\gamma(\hat{\psi}))$. We can also target other marginal quantities of interest by taking sample averages of $\gamma_{m,k}(\bar{A}_m,\bar{h}_m;\hat{\psi})$ in various subpopulations. For example, the sample mean of $\gamma_{1,2}(\bar{A}_1,\bar{h}_1;\hat{\psi})$ among units with $h_1=1$ and $A_1=0$ is a consistent estimator of the average immediate effect of indirect exposure among those indirectly but not directly treated at the second time point, marginalizing over direct exposure history. Similarly, the sample mean of $\gamma_{1,2}(\bar{a}_1,\bar{h}_1;\hat{\psi})$ among units with $a_1=1$ and $h_1=0$ is a consistent estimator of the average immediate effect of direct exposure among those directly but not indirectly treated at the second time point, marginalizing over indirect exposure history.

We can compare the estimator of the expected outcome had no unit ever been treated ($\hat{E}[\bar{Y}(\bar{0},\bar{0})]$) obtained from our DiD-SNMM modeling interference to an estimate of the same quantity from a DiD-SNMM that incorrectly assumes no interference. We specify a saturated DiD-SNMM under the incorrect assumption of no interference
\[
\gamma_{m,k}^{'}(\bar{a}_m) = E[Y_k(\bar{a}_m,\underbar{0})-Y_k(\bar{a}_{m-1},\underbar{0})|\bar{A}_m=\bar{a_m}]
\]
with parameters
 $\psi_{0,1}^{'}=\gamma^{'}_{0,1}(a_0=1)$, $\psi_{0,2}^{'}=\gamma^{'}_{0,2}(a_0=1)$, and $\psi_{1,2}^{'}=\gamma^{'}_{1,2}(\bar{a}_1=1)$. We compute estimator $\hat{\psi}^{'}$ as described in \citet{shahn2022structural}. We then estimate $E[Y_2(\bar{0})]$ by $\mathbf{P}_n \{Y_2-\psi_{1,2}^{'}A_1-\psi_{0,2}^{'}A_0\}$. In our simulation, the true value of $E[Y_2(\bar{0},\bar{0})]$ is 0.5. The mean estimate we obtained from our DiD-SNMM that ignored interference was 1.3 (sd=0.013) with 0\% coverage of bootstrap 95\% confidence intervals. (However, the bootstrap estimates of the standard error were unbiased with mean 0.013.) Estimates of $\hat{E}[\bar{Y}(\bar{0},\bar{0})]$ derived from the DiD-SNMM that modeled interference, by contrast, were unbiased, and bootstrap 95\% confidence intervals had nominal coverage. The lesson is that even if spillover effects are not directly of interest, failure to model them if they are present can lead to significant bias in estimates of other quantities.

\subsection{Cluster Interference}
We illustrate our approach in the cluster interference setting with a simulation of another two time-step data generating process. We define clusters comprising two units each such that spillover effects are contained within clusters (similar to the eyes application of \citet{he2015structural}). As in Section \ref{cluster}, we let $X_{i,t,j}$ denote the observation of arbitrary time-varying variable $X$ for unit $j$ of cluster $i$ at time $t$. The data generating process was as follows:
\begin{align*}
    &U_i\sim Bernoulli(0.5)\\
    &\begin{bmatrix}
        Y_{i,0,1}\\
        Y_{i,0,2}
    \end{bmatrix}\sim N(\begin{bmatrix}
        U_i\\
        U_i
    \end{bmatrix},\begin{bmatrix}
        0.1\text{      }0\\
        0\text{      }0.1
    \end{bmatrix})\\
    &\begin{bmatrix}
        Y_{i,1,1}(A_{i,0,1}=A_{i,0,2}=0)\\
        Y_{i,1,2}(A_{i,0,1}=A_{i,0,2}=0)
    \end{bmatrix}\sim N(\begin{bmatrix}
        U_i\\
        U_i
    \end{bmatrix},\begin{bmatrix}
        0.1\text{      }0\\
        0\text{      }0.1
    \end{bmatrix})\\
    &\begin{bmatrix}
        Y_{i,2,1}(\bar{A}_{i,1,1}=\bar{A}_{i,1,2}=\bar{\textbf{0}})\\
        Y_{i,2,2}(\bar{A}_{i,1,1}=\bar{A}_{i,1,2}=\bar{\textbf{0}})
    \end{bmatrix}\sim N(\begin{bmatrix}
        U_i\\
        U_i
    \end{bmatrix},\begin{bmatrix}
        0.1\text{      }0\\
        0\text{      }0.1
    \end{bmatrix})\\
    &A_{i,0,1},A_{i,0,2}\sim Bernoulli(0.3+0.2U_i)\\
    &\begin{bmatrix}
        Y_{i,1,1}\\
        Y_{i,1,2}
    \end{bmatrix}\sim N(\begin{bmatrix}
        Y_{i,1,1}(A_{i,0,1}=A_{i,0,2}=0)\\
        Y_{i,1,2}(A_{i,0,1}=A_{i,0,2}=0)
    \end{bmatrix}+\gamma_{01}(A_{i,0,1},A_{i,0,2}),\begin{bmatrix}
        0.1\text{      }0\\
        0\text{      }0.1
    \end{bmatrix})\\
    &A_{i,1,1}\sim (1-A_{i,0,1})Bernoulli(0.3 + 0.2U_i)\\
    &A_{i,1,2}\sim (1-A_{i,0,2})Bernoulli(0.3 + 0.2U_i)\\
    &\begin{bmatrix}
        Y_{i,1,1}\\
        Y_{i,1,2}
    \end{bmatrix}\sim N(\begin{bmatrix}
        Y_{i,2,1}(\bar{A}_{i,1,1}=\bar{A}_{i,1,2}=\bar{\textbf{0}})\\
        Y_{i,2,2}(\bar{A}_{i,1,1}=\bar{A}_{i,1,2}=\bar{\textbf{0}})
    \end{bmatrix} + \gamma_{0,2}(A_{i,0,1},A_{i,0,2}) + \gamma_{1,2}(\bar{A}_{i,1,1},\bar{A}_{i,1,2}), \begin{bmatrix}
        0.1\text{      }0\\
        0\text{      }0.1
    \end{bmatrix})
\end{align*}
with blip functions
\begin{align*}
    \undertilde\gamma_{0,1}(a_{01},a_{02})=\begin{bmatrix}
        \psi_{0,1}^1a_{01} + \psi_{0,1}^2a_{02} \\
         \psi_{0,1}^1a_{02} + \psi_{0,1}^2a_{01}
    \end{bmatrix}\\
    \undertilde\gamma_{0,2}(a_{01},a_{02})=\begin{bmatrix}
        \psi_{0,2}^1a_{01} + \psi_{0,2}^2a_{02} \\
         \psi_{0,2}^1a_{02} + \psi_{0,2}^2a_{01}
    \end{bmatrix}\\
    \undertilde\gamma_{1,2}(\bar{a}_{11},\bar{a}_{12})=\begin{bmatrix}
        \psi_{1,2}^1a_{11} + \psi_{1,2}^2a_{12} +
         \psi_{1,2}^3a_{11}(a_{02} + a_{12}) \\
         \psi_{1,2}^1a_{12} + \psi_{1,2}^2a_{11} +
         \psi_{1,2}^3a_{12}(a_{01} + a_{11})
    \end{bmatrix}
\end{align*}
and $(\psi_{0,1}^1=1,\psi_{0,1}^2=0.5,\psi_{0,2}^1=2,\psi_{0,2}^2=1,\psi_{1,2}^1=0.75,\psi_{1,2}^2=0.25,\psi_{1,2}^3=0.1)$.

The blip models impose parametric restrictions in this simulation example, unlike the network interference simulation. One restriction is that the effects of exposures on units in each cluster are symmetric. That is, the blip functions for effects on outcomes in unit $j$ have the same coefficients for direct and indirect exposures as the blip models for effects on outcomes in unit $j'$. For example, $\psi_{0,1}^1$ represents the immediate direct effect of exposure at time $0$ in both cluster units $j=1$ and $j=2$. Similarly, $\psi_{0,1}^2$ represents the immediate indirect effect of exposure at time $0$ in both cluster units $j=1$ and $j=2$. Another restriction is that there are no interactions between direct and indirect exposures in the blip models for effects of exposures at time $0$. However, effects of direct exposure at time $1$ depend on whether there was any indirect exposure through time $1$ via the interaction coefficient $\psi_{1,2}^3$. An advantage of SNMMs is the ability to exploit assumptions about effect heterogeneity via parametric blip model specification to gain efficiency as sample size constraints dictate. 

We simulated 500 data sets of $N=5,000$ clusters. We then proceeded to estimate $\psi$ nonparametrically by solving estimating equations (\ref{estimating_equations_cluster}) with plugged in saturated sample average estimators $\hat{E}_{\undertilde A_m|\bar{\undertilde{A}}_{m-1}}$ and $\hat{v}_m(k,\bar{\undertilde{a}}_{m-1};\undertilde\gamma(\psi))$ of the nuisance functions $E_{\undertilde A_m|\bar{\undertilde{A}}_{m-1}}$ and $v_m(k,\bar{\undertilde{a}}_{m-1};\undertilde\gamma(\psi))$. Standard errors were estimated using the sandwich estimator described in Section \ref{cluster_se}. All parameters of interest were approximately unbiasedly estimated, and 95\% confidence interval coverage was nominal.  

\begin{table}[t!]
\centering
\caption{Cluster interference simulation results. Standard errors and coverage computed using the sandwich variance estimator from Section \ref{cluster_se}}.
\begin{tabular}{lcccc}
\hline
\textbf{Estimand} & \textbf{True} & \textbf{Mean (sd)} & \textbf{SD(est)} & \textbf{95\% cov.} \\
\hline
$\psi^{1}_{0,1}$ & 1.00 & 1.00 (.003) & 0.00376 & 0.940 \\
$\psi^{2}_{0,1}$ & 0.50 & 0.50 (.002) & 0.00352 & 0.948 \\
$\psi^{1}_{0,2}$ & 2.00 & 2.00 (.003) & 0.00418 & 0.944 \\
$\psi^{2}_{0,2}$ & 1.00 & 1.00 (.003) & 0.00452 & 0.962 \\
$\psi^{1}_{1,2}$ & 0.75 & 0.75 (.006) & 0.00819 & 0.954 \\
$\psi^{2}_{1,2}$ & 0.25 & 0.25 (.004) & 0.00576 & 0.958 \\
$\psi^{3}_{1,2}$ & 0.007 & 0.10 (.014) & 0.00993 & 0.958 \\
\hline
\end{tabular}
\end{table}

\section{Real Data Application: Medicaid Expansion and Uninsurance}\label{sec:app}

In an analysis similar to the simulation in Section \ref{subsection:network_sim}, we estimated the direct and spillover effects of Medicaid expansion under the Affordable Care Act on county-level uninsurance rates in the population under 65 years of age. We defined a network interference mapping in which a county's indirect exposure was a binary indicator of whether any geographically bordering county was in a different state that expanded, i.e., $D_{i,m}=\phi_m(\textbf{A}_m,\theta_{i,m})=(A_{i,m},h_m=max(A_{j,m}))$ for $j\in\mathcal{N}_i$ with $\mathcal{N}_i$ denoting the neighbors of $i$. We looked at the three years 2013-2015. The first (and most) expansions occurred in 2014, and a few states expanded in 2015. 

It has been well established that Medicaid expansion reduced uninsurance in expansion states, which was its direct goal. However, one might wonder if there were also spillover effects. One hypothetical mechanism could be through heightened awareness of the importance of health insurance through outreach campaigns by providers in neighboring expansion states. 

We specified the blip models 
\begin{align}
     &\gamma_{0,k}(a_0,h_1) = \psi_1a_0 + \psi_2h_0 + \psi_3a_0(k-1) + \psi_4h_0(k-1) + \psi_5a_0h_0 +\psi_6a_0h_0(k-1)\\
     &\gamma_{2,2}(\bar{a}_1,\bar{h}_1) = \psi_7a_1 + \psi_8h_1 +   
    \psi_{9}a_1h_0 +\psi_{10}h_1a_0+\psi_{11}h_1h_0.
\end{align}
These are similar to the model specifications from Section \ref{subsection:network_sim}. The only difference is that we did not include terms $a_1h_1$ or $a_1h_0h_1$ because there were no counties in states that first expanded in 2015 and also adjacent to other states that first expanded in 2015. We estimated $\psi$ by solving estimating equations (\ref{estimating_equations}) with plugged in saturated sample average estimators $\hat{E}_{D_m|\bar{D}_{m-1}}$ and $\hat{v}_m(k,\bar{d}_{m-1};\gamma(\psi))$ of the nuisance functions $E_{D_m|\bar{D}_{m-1}}$ and $v_m(k,\bar{d}_{m-1};\gamma(\psi))$. We estimated standard errors using a spatial block bootstrap procedure \citep{lahiri2013resampling, sherman1996variance}. To create blocks, we laid a grid of hexagons over the map of the United States, each of width 75 km. Each hexagon then defined a block comprising all counties contained in it. We sampled these blocks randomly with replacement to create our bootstrap samples. 

Focusing on effects of 2014 expansions, Table \ref{tab:app} summarizes our results. We confirm the finding of direct effects of Medicaid expansion on uninsurance rates, but find no evidence of spillover effects. Our estimates of the contrasts $\gamma_{0,1}(a_0=1,h_0=0)$ and $\gamma_{0,1}(a_0=1,h_0=1)$ indicate that counties in states that expanded in 2014 on average reduced uninsurance rates in the same year by about $0.3$ percentage points, whether or not neighboring counties were also in states that expanded. Our estimates of the contrasts $\gamma_{0,2}(a_0=1,h_0=0)$ and $\gamma_{0,2}(a_0=1,h_0=1)$ indicate that counties in states that expanded in 2014 on average reduced uninsurance rates in 2015 by about $0.8$ percentage points, whether or not neighboring counties were also in states that expanded. Our estimates of $\gamma_{0,1}(a_0=0,h_0=1)$ and $\gamma_{0,2}(a_0=0,h_0=1)$ provide no evidence that neighboring an expansion state had any immediate or delayed impact on uninsurance rates in counties that did not expand. However, the confidence intervals do not rule out small effects. Given our prior beliefs about the likely sizes of the effects estimated here, we interpret our results as evidence for the soundness of our method.

\begin{table}[t!]
\centering
\caption{Estimates and spatial bootstrap standard errors of direct and spillover effects of Medicaid expansion on county level uninsurance rates under a geographic proximity network interference structure. Units are percentage points.}
\begin{tabular}{lccc}
\hline
\textbf{Estimand} & \textbf{Estimate (95\% CI)}\\
\hline
$\gamma_{0,1}(a_0=1,h_0=0)$                     & -0.31 (-0.40,-0.22) \\
$\gamma_{0,1}(a_0=1,h_0=1)$                     & -0.32 (-0.46,-0.18) \\
$\gamma_{0,1}(a_0=0,h_0=1)$                     &  0.05 (-0.10,0.20)\\
$\gamma_{0,2}(a_0=1,h_0=0)$                     & -0.78 (-0.93,-0.63) \\
$\gamma_{0,2}(a_0=1,h_0=1)$                     & -0.77 (-0.99,-0.54)  \\
$\gamma_{0,2}(a_0=0,h_0=1)$                     & -0.07 (-0.31,0.17) \\
\hline
\end{tabular}\label{tab:app}
\end{table}

\section{Conclusion}\label{sec:conclusion}
The main contribution of this paper consists of the observation that the flexibility of SNMMs to model effects of repeated and multidimensional exposures naturally lends itself to jointly modeling effects of direct and indirect or spillover exposures. Previous results on identification and estimation of SNMMs \citep{shahn2022structural} under parallel trends assumptions immediately extend to settings with network interference and a known exposure mapping or with cluster interference. Applying SNMMs in these settings expands what can be learned about effects of exposures in the presence of spillovers under parallel trends assumptions beyond what was possible with alternative DiD approaches. In our discussions of identification and simulations, we provided an indication of the sorts of estimands that might be targeted via SNMMs. Depending on the application, different byproducts or parameters of the blip functions might be of substantive interest. One of the key features of SNMMs that we did not highlight in the simulations is the ability to naturally accommodate continuous treatments. Another is the ability to model heterogeneity of direct and spillover effects as a function of time-varying covariates.

In the paper, we presented doubly robust estimating equations and described an estimation procedure based on plugging in parametric nuisance models. However, if data-adaptive machine learning methods are employed for the nuisance models, a cross-fitting procedure \citep{chernozhukov2018double} can be used for estimation to retain regular asymptotically linear estimators with valid confidence intervals in the cluster interference setting. \citet{shahn2022structural} provides some guidance on implementation. In the network interference setting, the theory of influence functions on which the cross-fitting procedure is based is less well understood, and it is unclear what impact cross-fitting would have on the operating characteristics of an estimator employing machine learning to estimate the nuisance models. \citet{jetsupphasuk2025estimating} discuss how and when their standard error estimator allows for double/debiased machine learning estimation procedures. 

\bibliography{references}

\end{document}